\documentclass{webofc}
\usepackage[varg]{txfonts}   
\usepackage[makeroom]{cancel}

\newcommand{\tr}{\mathop{\mbox{tr}}\nolimits}

\begin{document}
\title{DIS structure functions in the NLO approximation of the Parton Reggeization Approach}
%
%

\author{\firstname{Maxim} \lastname{Nefedov}\inst{1}\fnsep\thanks{\email{nefedovma@gmail.com}} \and
        \firstname{Vladimir} \lastname{Saleev}\inst{1}\fnsep\thanks{\email{saleev@samsu.ru}} }

\institute{Samara National Research University, Moskovskoe Shosse, 34, 443086, Samara, Russia
          }

\abstract{%
  The main ideas of the NLO calculations in Parton Reggeization Approach are illustrated on the example of the simplest NLO subprocess, which contributes to DIS: $\gamma^\star+R\to q + \bar{q}$. The double counting with the LO contribution $\gamma^\star + Q \to q$ is resolved. The problem of matching of the NLO results for single-scale observables in PRA on the corresponding NLO results in Collinear Parton Model is considered. In the developed framework, the usual NLO PDFs in the $\overline{MS}$-scheme can be consistently used as the collinear input for the NLO calculation in PRA.
}
\maketitle
\section{Introduction}
\label{intro}

  Multiscale and correlational observables, related with the hard processes in the hadronic or lepton-hadronic collisions, still present a challenge for the state-of-the-art calculational methods in perturbative QCD. By such observables we mean, for example, angular correlations of pairs of vector bosons, jets or reconstructed hadrons, or observables related with the polarization of virtual photon/Z-boson in the Drell-Yan process. All these observables are highly sencitive to QCD radiation over wide range of scales -- from soft and collinear to the hard emissions with transverse momenta of the order of the hard scale. Parton showers (PS) are most accurate in modelling of the soft and collinear emissions, while fixed-order calculations in QCD demonstrate good convergence only for single-scale quantities. The aim of the Parton Reggeization Approach (PRA) is to improve our understanding of the transition between soft/collinear and hard regimes, using the information from the Multi-Regge limit of scattering amplitudes in QCD, and thus to reduce the gap between fixed-order and PS Monte-Carlo (MC) techniques.  
  
  In the present contribution we continue the development of formalism of the Next-to-Leading Order (NLO) calculations in PRA, concentrating on the classic example of {\it single-scale} observable: $F_2(x_B,Q^2)$-structure function of the electron-proton Deep Inelastic Scattering (DIS). For the single-scale observables, the NLO results of PRA should be consistent with the corresponding NLO results in the Collinear Parton Model (CPM). We exploit this relation to fix the formalism of NLO calculations in PRA in such a way, that the Parton Distribution Functions (PDFs) from NLO global fits, which are conventionally defined in the $\overline{MS}$-scheme, can be used as an input to the NLO calculations in PRA.
  
  The discussion of loop corrections in PRA has been started by us in the Ref.~\cite{Loop_proc}. In the present contribution we concentrate exclusively on tree-level corrections, and issues related with the double-counting and scheme-dependence of PDFs. The present paper has the following structure. In the Sec.~\ref{sec-1} the LO-formalism of PRA is reviewed and the main ideas beyond the NLO calcualtions in PRA are formulated. In the Sec.~\ref{sec-2} the first $O(\alpha_s)$ contribution, which comes from the dependence of the LO hard-scattering coefficient (HSC) in PRA on the transverse momentum of initial-state parton is isolated. In the Sec.~\ref{sec-3} the contribution of the NLO subprocess $\gamma^\star + R \to q + \bar{q}$ and the corresponding double-counting subtraction is introduced. Finally, in the Sec.~\ref{sec-4} the formulated scheme of NLO calculations is tested numerically.      

\section{DIS at LO and NLO framework in PRA}
\label{sec-1}

\begin{figure}
\begin{tabular}{cc}
\hspace{-4mm}\includegraphics[width=0.45\textwidth]{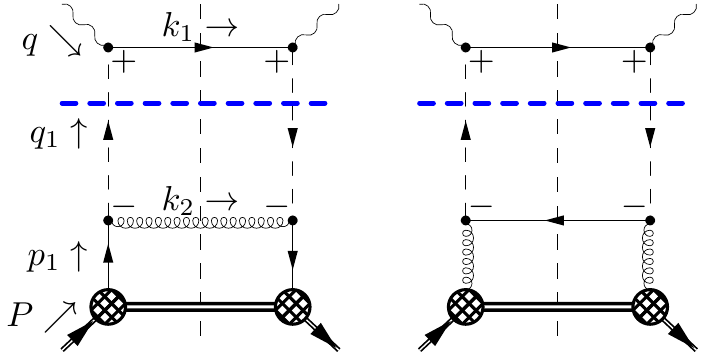} &
\includegraphics[width=0.54\textwidth]{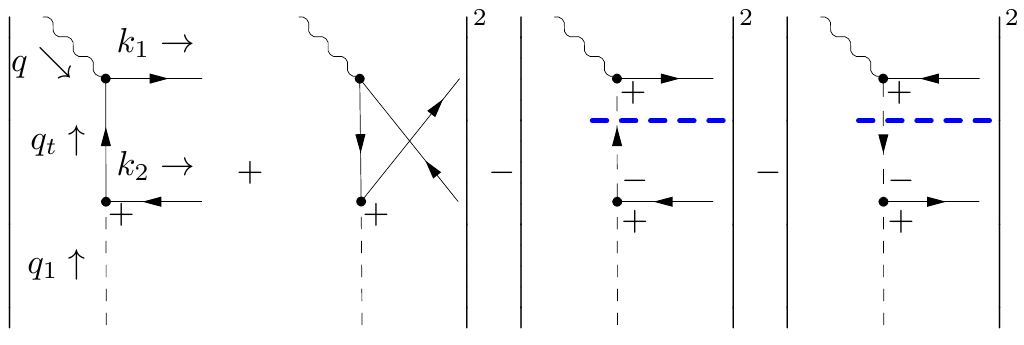} \\
(a) & (b)
\end{tabular}
\caption{Panel (a): contributions to the DIS structure functions in the LO of PRA. Panel (b): Contribution $\gamma^\star + R \to q + \bar{q}$ and double-counting subtraction terms. The horizontal thick dashed line denotes the mMRK-approximation, i.e. the ``small'' $q_1^-$ and $q_t^-$ light-cone components of $t$-channel momenta do not propagate to the upper part of the graph.\label{fig:LO-NLO}}
\end{figure}

  We will consider the process of lepton-proton DIS, which is traditionally described as a process of desintegration of the proton with the four-momentum $P_\mu$ (we put $P^2=0$, since $E_p\gg M_p$) by the virtual photon with the four-momentum $q_\mu$ ($q^2=-Q^2<0$). We will work in the center-of-mass frame of the proton and virtual photon, where the Sudakov(light-cone) components\footnote{We use the following Sudakov decomposition for the four-momentum $k$: $k^\mu=(n_+^\mu k^- + n_-^\mu k^+)/2 + k_T^\mu$, $n_{\pm}^\mu = (n^{\pm})^\mu=(1,0,0,\mp 1)^{\mu}$, $n_\pm k_T=0$, $k^\pm=k_{\pm}=(n^\pm k)$, so that $k^2=k_+k^- - {\bf k}_T^2$, $P^+\neq 0$ and $P^-=0$. } of momentum of the photon are:
  \[
  q^+=-x_BP^+,\ \ q^-=\frac{Q^2}{x_BP^+},\ \ {\bf q}_{T}=0,
  \]
  where $x_B=Q^2/(2qP)$ is the usual Bjorken variable. Inclusive DIS is fully described by the hadronic tensor, with the following standard parametrization in terms of structure functions~\cite{IFL, Collins}:
  \begin{equation}
  W^{\mu\nu}=\left(-g_{\mu\nu}+\frac{q_\mu q_\nu}{q^2}\right)F_1(x_B,Q^2)+\left(P_\mu-\frac{q_\mu (Pq)}{q^2}\right)\left(P_\nu-\frac{q_\nu (Pq)}{q^2}\right) \frac{F_2(x_B,Q^2)}{(Pq)}, \label{eq:Wmn-def}
  \end{equation}
  where the function $F_1$ is related with the commonly used structure function $F_L(x_B,Q^2)$ as follows: $F_1=(F_2-F_L)/({2x_B})$. 
  
  In the CPM, the hadronic tensor is represented as a convolution of the partonic tensor of hard interaction of the virtual photon with the partonic target and the PDF $f_i(x,\mu^2)$. To derive the factorization formula of PRA for DIS, we will start with the specific NLO CPM hard subprocesses:
  \begin{eqnarray}
 \gamma^\star(q) + q(p_1) \to  q(k_1) + g(k_2), \label{subp:gam_q-q_g} \\
 \gamma^\star(q) + g(p_1) \to  q(k_1) + \bar{q}(k_2). \label{subp:gam_g-q_q}
  \end{eqnarray}   
  
  Following the general scheme of LO PRA calculations, which is described in more detail in the Refs.~\cite{Proc_HQ, BB-correlations}, we write down the modified MRK (mMRK) approximation for the partonic tensor of processes (\ref{subp:gam_q-q_g}) and (\ref{subp:gam_g-q_q}):
  \begin{equation}
  \hat{w}_{q/g}^{\mu\nu} = \frac{2g_s^2}{(-q_1^2)} \frac{P_{q(q/g)}(\tilde{z})}{\tilde{z}} \hat{w}^{\mu\nu}_{LO}, \label{eq:mMRK_LO}
  \end{equation}
  where $q_1=p_1-k_2$, $g_s=\sqrt{4\pi \alpha_s}$ is the coupling contant of strong interaction, $\tilde{z}=q_1^+/p_1^+$, $P_{q(q/g)}(z)$ denotes the (non-regularized) DGLAP splitting functions: $P_{qq}(z)=C_F (1+z^2)/(1-z)$ or $P_{qg}(z)=T_R\left[z^2+(1-z)^2\right]$ for process (\ref{subp:gam_q-q_g}) and (\ref{subp:gam_g-q_q}) respectively, $C_F=(N_c^2-1)/(2N_c)$, $T_R=1/2$, $N_c=3$, and $\hat{w}^{\mu\nu}_{LO}$ is the LO partonic tensor in PRA. The diagrammatic representation of Eq.~(\ref{eq:mMRK_LO}) is given in the Fig.~\ref{fig:LO-NLO}(a). The dashed lines (dashed lines with arrow) in the Fig.~\ref{fig:LO-NLO} denote Reggeized gluons($R$) (quarks -- $Q$). By definition of the mMRK-approximation, the $q_1^-$ light-cone component of the momentum $q_1$ does not propagate to the $\gamma Qq$-scattering vertex, so that the LO partonic tensor is given by the hard subprocess $\gamma^\star(q)+Q(\tilde{q}_1)\to q(k_1)$, where $\tilde{q}_1^\mu=q_1^+ n_-^\mu /2 + q_{T1}^\mu$. The partonic tensor of this subprocess reads:
\begin{equation}
 \hat{w}^{\mu\nu}_{LO}=\frac{e_q^2}{2}\tr\left[ \hat{k}_1\Gamma^\mu(\tilde{q}_1,q)\left( \frac{q_1^+}{2} \hat{n}_- \right) \Gamma^\nu(\tilde{q}_1,q) \right], \label{eq:what-LO}
\end{equation}  
 where $\Gamma^\mu(q_1,k)=\gamma^\mu+\hat{q}_1(n_-^\mu)/k^-$ is the gauge-invariant effective vertex of $\gamma Qq$-interaction (Fadin-Sherman vertex~\cite{Fadin-Sherman, ReggeProofLL}), and $e_q$ is the electric charge of the quark in units of the electron charge. 
 
 Substituting the results (\ref{eq:mMRK_LO}) and (\ref{eq:what-LO}) to the factorization formula of the CPM and projecting-out the $F_2$ structure function, one can rewrite the structure function in the {\it $k_T$-factorized form}:
 \begin{equation}
 F_{2q}^{(LO\ PRA)}(x_B, Q^2) = \int\limits_{x_B}^1 \frac{dx_1}{x_1} \int\limits_0^\infty dt_1\ \left[ \tilde{\Phi}_q(x_1,t_1,Q^2) + \tilde{\Phi}_{\bar{q}}(x_1,t_1,Q^2) \right] \cdot C_{2}^{(0)}\left(\frac{x_B}{x_1}, \frac{t_1}{Q^2}\right), \label{eq:factor}
 \end{equation}  
 where the HSC $C_{2}^{(0)}$ takes the form:
 \begin{equation}
 C_{2}^{(0)}\left(z, \frac{t_1}{Q^2}\right) = e_q^2\cdot  z \delta\left(  \left(1+\frac{t_1}{Q^2}\right)z - 1 \right),\label{eq:LO_HSC}
 \end{equation}  
 and the tree-level ``unintegrated PDF'' (unPDF) is defined as:
 \begin{equation}
 \tilde{\Phi}_i(x,t,\mu^2) = \frac{1}{t} \frac{\alpha_s}{2\pi} \sum\limits_{j=q,\bar{q},g} \int\limits_x^1 d\tilde{z}\ P_{ij}(\tilde{z})\cdot \frac{x}{\tilde{z}}f_j\left(\frac{x}{\tilde{z}} ,\mu^2 \right). \label{eq:tree-unPDF}
 \end{equation}
 
 The $F_L$ structure function is equal to zero in the LO of PRA, due to the properties of the Fadin-Sherman vertex.
 
 Integration over $t_1$ in the Eq.~(\ref{eq:factor}) with ``unPDF'' (\ref{eq:tree-unPDF}) is logarithmically divergent at $t_1\to 0$ and at $\tilde{z}\to 1$ (for the case of diagonal splitting fuctions). The latter divergence is regularized in PRA by the following cutoff on $\tilde{z}$-variable:
 \begin{equation}
 \tilde{z}<1-\Delta_{KMR}(t,\mu^2),\label{eq:KMR-cut}
 \end{equation}
 where the Kimber-Martin-Ryskin~\cite{KMR} cutoff function $\Delta_{KMR}(t,\mu^2)=\sqrt{t}/(\sqrt{\mu^2}+\sqrt{t})$ is given by the condition of ordering in rapidity between the particles produced in the hard process and the parton emitted on the last step of the ISR parton cascade in the kinematics of $pp$-collisions\footnote{In the case of DIS, the rapidity ordering leads to the weaker condition on $\tilde{z}$: $\tilde{z}<1-t_1/(Q^2+2t_1)$, but we will keep KMR condition for consistency.}. 
 
 Collinear divergence at $t_1\to 0$ in (\ref{eq:tree-unPDF}) is regularized by the introduction of Sudakov formfactor, which resums the doubly-logarithmic corrections $\sim \log^2(t_1/\mu^2)$ in LLA, analogously to the treatment of this divergence in standard PS algorithms~\cite{MC-rev}. The exact expression of this formfactor is dictated by the DGLAP evolution for PDFs $f_i(x, \mu^2)$ and the following normalization condition for unPDFs:
 \begin{equation}
 \int\limits_0^{\mu^2} dt\ \Phi_i(x,t,\mu^2) = xf_i(x,\mu^2), \label{eq:unPDF-norm}
 \end{equation}  
 which, as it will be explained in the Sec.~\ref{sec-2}, is equivelent to the normalization of single-scale observables on the corresponding LO CPM results up to NLO terms in $\alpha_s$ and power-supressed corrections. Condition (\ref{eq:unPDF-norm}) is obviously fulfilled by the following {\it derivative form} of unPDF:
 \begin{equation}
 \Phi_i(x,t,\mu^2) = \frac{\partial}{\partial t}\left[ T_i(t,\mu^2,x) \cdot xf_i(x,t) \right], \label{eq:unPDF-der}
 \end{equation}
 where $T_i$ is the Sudakov formfactor with the boundary conditions\footnote{The first boundary condition is actually nonperturbative and introduced only to simplify the presentation of the analytic arguments. In our numerical calculations, the unPDF $t\cdot\Phi_i(x,t,\mu^2)$ is defined by quadratic polynomial for $t<Q_0^2=1$ GeV$^2$, which is adjusted to ensure continuity and the validity of Eq.~\ref{eq:unPDF-norm}. Other nonperturbative definitions are also possible.} $T_i(0,\mu^2,x)=0$ and $T_i(\mu^2,\mu^2,x)=1$. Eq.~\ref{eq:unPDF-der} can be shown to coincide exactly with the following {\it integral form} of unPDF:
 \begin{equation}
 \Phi_i(x,t,\mu^2)= \frac{T_i(t,\mu^2,x)}{t} \frac{\alpha_s(t)}{2\pi} \sum\limits_{j=q,\bar{q},g} \int\limits_{x}^1 d\tilde{z}\ P_{ij}(\tilde{z})\cdot \frac{x}{\tilde{z}} f_j\left(\frac{x}{\tilde{z}}, t \right) \cdot \Theta(\tilde{z},t,\mu^2)  , \label{eq:unPDF-int}
 \end{equation}   
 where $\Theta(\tilde{z},t,\mu^2)=\theta\left( 1-\Delta_{KMR}(t,\mu^2) -\tilde{z} \right)$, and the Sudakov formfactor is given by:
 \begin{eqnarray}
 &T_i(t,\mu^2,x)=\exp \left[ - \int\limits_t^{\mu^2} \frac{dt'}{t'} \frac{\alpha_s(t')}{2\pi} \left( \tau_i + \Delta\tau_i \right) \right],\ \ \ \tau_i = \sum\limits_j \int\limits_0^{1} d\tilde{z}\ \Theta(\tilde{z},t',\mu^2) \cdot \tilde{z} P_{ji}(\tilde{z}), \nonumber\\
 & \Delta\tau_i = \sum\limits_j \int\limits_0^1 d\tilde{z} \ \left(1-\Theta(\tilde{z},t',\mu^2) \right)
     \cdot \left[ \tilde{z}P_{ji}(\tilde{z}) - {\frac{\frac{x}{\tilde{z}} f_j\left(\frac{x}{\tilde{z}},t' \right)}{x f_i(x,t')}} P_{ij}(\tilde{z}) \cdot \theta(\tilde{z}-x)  \right]. \label{eq:Sudakov}
 \end{eqnarray}
 
 The integral form of unPDF~(\ref{eq:unPDF-int}) reproduces Eq.~\ref{eq:tree-unPDF} when $t\sim \mu^2$, in this way it is connected with the derivation of factorization formula (\ref{eq:factor}). Another important property of the Eqns.~\ref{eq:unPDF-int} -- \ref{eq:Sudakov} is that they are equivalent to Eq.~\ref{eq:unPDF-der} in any order in $\alpha_s$ for the evolution kernels $P_{ij}(z)$ if the PDFs $f_i(x,\mu^2)$ are defined in the corresponding subtraction scheme, e. g. both in the $\overline{MS}$-scheme.
 
 The LO framework of PRA has been formulated above, on the example of DIS process. Extension to the NLO would require the definition of subtraction scheme for double counting and {\it rapidity divergences} (see Ref.~\cite{Loop_proc}) in the real and virtual NLO corrections to the HSC, as well as the definition of NLO unPDFs. As the main guidance to resolve all these issues we will employ the {\it physical normalization condition} (PNC), which simply states, that for a given set of collinear PDFs, defined in the $\overline{MS}$ scheme, the NLO PRA calculation of any single-scale observable with the scale $Q^2$ should reproduce the corresponding NLO CPM result up to corrections, which are formally of higher order in $\alpha_s(Q^2)$ or supressed by the negative power of $Q^2$. Validity of this condition in QCD is guaranteed by the standard collinear factorization theorems~\cite{Collins}. To formulate PNC, we coose the {\it structure-function of DIS on a specified quark flavor} -- $F_{2q}(x_B,Q^2)$ as a benchmark single-scale gauge-invariant quantity, which is well-defined in any order of QCD perturbation theory. In the present paper we will study the contributions to PNC, which are related to the NLO PRA subprocess:
 \begin{equation}
 \gamma^\star(q) + R(q_1)\to q(k_1) + \bar{q}(k_2), \label{eq:NLO-subp}
 \end{equation}
 and we will match them to the corresponding $O(\alpha_s)$ contribution of gluon PDF in the NLO of CPM for $F_{2q}(x_B,Q^2)$. The NLO CPM test function, which is to be compared with our NLO PRA results, reads:
 \begin{equation}
 \tilde{F}_{2q}^{(NLO\ CPM)}(x_B, Q^2)=(e_q^2x_B)\left\{ f^{\overline{MS}}_q(x_B,Q^2) + f^{\overline{MS}}_{\bar{q}}(x_B,Q^2)  + 2 \left[C_{2g}^{\overline{MS}} \otimes f^{\overline{MS}}_g\right] (x_B,Q^2)  \right\}, \label{eq:CPM-test}
 \end{equation}    
 where the Mellin convolution is defined as: $\left[ C \otimes f \right](x,Q^2)= \int \limits_x^1 \frac{dz}{z} C(x/z)f(z,Q^2)$ and the textbook~\cite{Collins} result for NLO HSC in $\overline{MS}$-scheme is: $
 C^{\overline{MS}}_{2g}(z)= T_R \frac{\alpha_s(Q^2)}{2\pi}  \left[ \left(z^2+(1-z)^2\right) \log \left( \frac{1-z}{z}\right) -8z^2+8z-1 \right]$.

\section{NLO contribution from $k_T$-factorization}
\label{sec-2}

 In this section we will study the effects of nontrivial dependence of the LO HSC (\ref{eq:LO_HSC}) on $t_1$. Substituting the derivative form of unPDF to the factorization formula, one obtains:
 \begin{equation}
 F_{2q}^{(LO\ PRA)}(x_B,Q^2) = \int \limits_{x_B}^1 \frac{dx_1}{x_1} \int\limits_0^\infty dt_1\ \frac{\partial}{\partial t_1} \left[ T_q(t,Q^2,x_1)\cdot x_1 f_q(x_1,t) \right] \cdot C_2^{(0)}\left( \frac{x_B}{x_1}, \frac{t_1}{Q^2} \right) + {\rm c.c.}, \label{eq:F2q_LO_1}
 \end{equation}  
where ``${\rm c.c.}$'' denotes the antiquark term. Since the unPDFs decrease like a negative power of $t_1$ for $t_1>Q^2$ and the HSC behaves as $1/t_1$ for $t_1\gg Q^2$, one can conclude, that the contribution to the integral from $t_1>Q^2$ is supressed by the negative power of $Q^2$, and therefore can be neglected, which is denoted by the $(\simeq)$ sign in the following equations. With the help of integration by parts, the residual integral over the range $0\leq t_1\leq Q^2$ can be decomposed as follows:
\begin{equation}
F_{2q}^{(LO\ PRA)}(x_B,Q^2) \simeq e_q^2x_B f_q(x_B,Q^2) + \Delta F_{2q}^{(f)}(x_B,Q^2) + \Delta F_{2q}^{(T)}(x_B,Q^2)+{\rm c. c.}, \label{eq:F2q_LO_dec}
\end{equation}
where the first term is the usual CPM expression, and the corrections, which originate from the dependence of $C_2^{(0)}$ on $t_1$, have the following form:
\begin{eqnarray}
& \Delta F_{2q}^{(f)}(x_B,Q^2) =  \int \limits_0^1 dx_1\ \int \limits_0^{Q^2} dt_1\  T_q(t_1,Q^2,x_1) \left[ f_q(x_1,Q^2) - f_q(x_1,t_1) \right]\cdot \frac{\partial}{\partial t_1} C_{2}^{(0)} \left(\frac{x_B}{x_1}  , \frac{t_1}{Q^2} \right), \label{eq:DF-f} \\
& \Delta F_{2q}^{(T)}(x_B,Q^2) = \int \limits_0^1 dx_1 \ f_q(x_1,Q^2)  \int \limits_0^{Q^2} dt_1 \left[1- T_q(t_1,Q^2, x_1) \right] \cdot \frac{\partial}{\partial t_1} C_{2}^{(0)} \left(\frac{x_B}{x_1} , \frac{t_1}{Q^2} \right) . \label{eq:DF-T}
\end{eqnarray}

 Expressions in the square brackets in Eqns.~\ref{eq:DF-f} and \ref{eq:DF-T} are of $O(\alpha_s(Q^2))$, since both the running of PDFs with the scale and difference of Sudakov formfactor from unity are perturbative effects. Therefore both corrections $\Delta F_{2q}^{(f)}$ and $\Delta F_{2q}^{(T)}$ are of the NLO in $\alpha_s$, and the condition (\ref{eq:unPDF-norm}) guarantees the physical normalization at LO, as it was stated in the Sec.~\ref{sec-1}. 
 
 To compute $\Delta F_{2q}^{(f)}$, one taylor-expands $f_q(x,Q^2)-f_q(x,t_1)$ in $\alpha_s(Q^2)\log(t_1/Q^2)$ using the DGLAP equations:
 \begin{equation}
 f_q(x,Q^2)-f_q(x,t_1) = - \frac{\alpha_s(Q^2)}{2\pi}\log\left(\frac{t_1}{Q^2}\right) \left\{ \left[ P_{qq} \otimes f_q \right](x,Q^2) +  \left[ P_{qg} \otimes f_g \right](x,Q^2) \right\} + O(\alpha_s^2(Q^2)). \label{eq:f-f}
 \end{equation}
  Since in the present paper we deal with the gluon-induced subprocess (\ref{eq:NLO-subp}), we will be interested only in the contribution, which contains the gluon PDF. Contribution to $\Delta F_{2q}^{(f)}$  which contains quark PDFs, as well as the $\Delta F_{2q}^{(T)}$ should be considered together with the $\gamma^\star + Q \to q + g$ subprocess and loop correction to the $\gamma^\star + Q \to q$ subprocess respectively. To compute the $O(\alpha_s)$ contribution to $\Delta F_{2q}^{(f_g)}$, terms of higher order in $\alpha_s$ in (\ref{eq:f-f}) and the Sudakov formfactor in the Eq.~\ref{eq:DF-f} can be omitted altogether. Applying integration by parts to the integral over $t_1$ in (\ref{eq:DF-f}), one obtains:
  \begin{eqnarray}
\hspace{-1.1cm} & \Delta F_{2q}^{(f_g)} = (e_q^2 x_B) \frac{\alpha_s(Q^2)}{2\pi} \int \limits_0^1 \frac{dx_1}{x_1} f_g(x_1,Q^2) \left\{ - P_{qg}\left(\frac{x_B}{x_1}\right) \log\left(\frac{Q^2}{\lambda^2}\right) +  \int\limits_{\lambda^2}^{Q^2} \frac{dt_1}{t_1}\int \limits_{x_B}^{x_1} dz\ P_{qg}\left(\frac{z}{x_1}\right)\delta\left(z - x_B \frac{Q^2+t_1}{Q^2} \right)  \right\}.\ \ \label{eq:DF_fg_1}
  \end{eqnarray}  
  The spurious collinear divergence is regulated by the cutoff $\lambda^2$ in Eq.~\ref{eq:DF_fg_1}, and it cancels between two terms in curly brackets. Taking the integrals over $z$ and $t_1$ one finds, that $\Delta F_{2q}^{(f_g)}(x_B,Q^2)=(e_q^2 x_B) \left[ \Delta C_{qg}^{(k_T)} \otimes f_g \right](x_B,Q^2)$, where:
  \begin{equation}
  \Delta C_{qg}^{(k_T)}(z) = T_R \frac{\alpha_s(Q^2)}{2\pi}\left[ \xi z \left((4+\xi)z-2 \right) + \left(z^2 + (1-z)^2\right) \log \xi \right],
\end{equation}   
and $\xi = \min \left(1,(1-z)/z\right)$. In such a way, the $O(\alpha_s)$ contributions, which arize from the $t_1$-dependence of LO HSC, can be calculated explicitly and taken into account in the PNC.

\section{Gluon-induced NLO contribution}
\label{sec-3}

Contribution of the subprocess (\ref{eq:NLO-subp}) to the HSC is depicted by the first term of diagrammatic expression in the Fig.~\ref{fig:LO-NLO}(b) and  explicitly depends on ${\bf q}_{T1}$. This contribution can be written down using Feynman Rules of Lipatov's theory, see e. g. Ref.~\cite{BB-correlations} and references therein. It is convenient work in terms of kinematic variables:
\begin{equation}
z = \frac{x_B}{x_1},\ \tilde{z}=\frac{q_1^+-k_2^+}{q_1^+},
\end{equation}
and express the transverse momentum ${\bf k}_{T1}$ in terms of ${\bf q}_{T1}$ and a new vector ${\bf k}_T$:
\begin{equation}
{\bf k}_{T1}={\bf q}_{T1} \frac{\tilde{z}-z}{1-z} + {\bf k}_{T}.
\end{equation}

  The square of ${\bf k}_T$ is fixed by the condition $k_1^2=0$:
  \begin{equation}
  {\bf k}_T^2=\frac{(\tilde{z}-z)(1-\tilde{z})}{1-z} \left[ \frac{Q^2}{z} - \frac{t_1}{1-z} \right],
  \end{equation}
  and we will integrate over azimuthal angle $\psi$ between ${\bf k}_{T}$ and ${\bf q}_{T1}$. In this notation, contribution of the subprocess (\ref{eq:NLO-subp}) to the $F_{2q}$ structure function takes the form:
  \begin{equation}
  F_{2q}^{(1,g)}(x_B,Q^2) = (e_q^2 x_B) \int \limits_{x_B}^1 \frac{dx_1}{x_1^2} \int\limits_0^{t_1^{\max}} dt_1\ \Phi_g(x_1,t_1,Q^2)\cdot C_{2}^{(1,g)}\left(\frac{x_B}{x_1},\frac{t_1}{Q^2} \right), \label{eq:F2q_1g}
  \end{equation}
  where the upper limit $t_1^{\max}=Q^2(1-z)/z$ follows from the condition $(k_1+k_2)^2>0$, 
  \begin{equation}
  C_{2\ NS}^{(1,g)}\left( z, \frac{t_1}{Q^2} \right) = \frac{\alpha_s(Q^2)}{2\pi} \int\limits_z^1 \frac{ d\tilde{z}}{(1-z)} \int\limits_0^{2\pi}  \frac{d\psi}{2\pi}\ {\cal C}^{(1,g)}_{2} \left(z, \tilde{z}, t_1 ,{\bf k}_{T1}^2, Q^2 \right), \label{eq:NLO-HSC_1}
  \end{equation}
  where subscript $NS$ denotes, that no double-counting subtractions has been performed yet, and
\begin{eqnarray*}
   {\cal C}_2^{(1,g)}&=& \frac{T_R}{t_1 \left(Q^2 (\tilde{z}-1) (z-\tilde{z})+{\bf k}_{T1}^2 z^2\right)^2 \left(Q^2 \tilde{z}
   (z-\tilde{z})+{\bf k}_{T1}^2 z^2\right)^2} \nonumber \\
    &\times & \left\{ Q^{10} (\tilde{z}-1)^2 (z-\tilde{z})^4+Q^8 (\tilde{z}-1) (z-\tilde{z})^3 (2 {\bf k}_{T1}^2 z
   (z+3 \tilde{z}-4)+t_1 \tilde{z} (z-\tilde{z}))\right.  \nonumber\\
   &+& Q^6 {\bf k}_{T1}^2 z (z-\tilde{z})^2 \left({\bf k}_{T1}^2 z
   \left(z^2+2 z (6 \tilde{z}-7)+6 (\tilde{z}-4) \tilde{z}+19\right) \right. \\
   &+& \left. t_1 (z-\tilde{z}) (z (2 \tilde{z}-1)+2
   (\tilde{z}-1) \tilde{z})\right)  \nonumber \\
   &+&  Q^4 {\bf k}_{T1}^4 z^2 (z-\tilde{z}) \left(6 {\bf k}_{T1}^2 (z-1) z (z+2
   \tilde{z}-3)+t_1 (z-\tilde{z}) \left(z^2+z (4 \tilde{z}-2)+2 (\tilde{z}-1) \tilde{z}\right)\right)  \nonumber \\
   &+&\left.  2 Q^2
   {\bf k}_{T1}^6 z^4 \left(3 {\bf k}_{T1}^2 (z-1)^2+t_1 (z-\tilde{z}) (z+2 \tilde{z}-1)\right)+2 {\bf k}_{T1}^8 t_1 z^6 \right\} .
\end{eqnarray*}

  Contribution of the subprocess (\ref{eq:NLO-subp}) clearly contains double-counting with the contribution of the second diagram in the Fig.~\ref{fig:LO-NLO}(a). To resolve this problem, we construct the corresponding mMRK double-counting subtraction (DCS) terms, which are depicted as the second and third terms of diagrammatic expression in the Fig.~\ref{fig:LO-NLO}(b). In these DCS terms one should implement the same kinematic approximations, as in the LO diagrams of the Fig.~\ref{fig:LO-NLO}(a), i. e. one have to neglect the $q_t^-$ momentum component on the entrance to the $\gamma Qq$-scattering vertex. This operation can be consistently performed by the following substitution: 
  \begin{eqnarray*}
&  \delta(q_1^+ + q^+ - k_1^+ - k_2^+)\delta(q^- - k_1^- - k_2^-) \to \int dq_t^+ dq_t^-\ \delta(q_1^+ - k_2^+ - q_t^+)\delta(k_2^- + q_t^-)\\
& \times  \delta(q^++q_t^+ - k_1^+) \delta(q^- + \xcancel{q_t^-} - k_1^-).
  \end{eqnarray*}
  
 In mMRK kinematics there is no need to introduce the vector ${\bf k}_T$, since the square of ${\bf k}_{T1}$ is expressed simply as:
  \begin{equation}
  {\bf k}_{\rm T1(mMRK)}^2 = Q^2 \frac{\tilde{z}-z}{z},
  \end{equation}
  and we will integrate over the angle $\phi$ between ${\bf k}_{T1}$ and ${\bf q}_{T1}$ instead of $\psi$. Also the factor $(1-z)$ in denominator of the Eq.~\ref{eq:NLO-HSC_1} is replaced by $(1-\tilde{z})$.
  
   Expression for the integrand of the $t$-channel DCS term reads:
  \begin{equation}
  \Delta {\cal C}^{(t)}_{2} = \frac{P_{QR}(\tilde{z},{\bf q}_{T1}, {\bf k}_{T1})}{\tilde{z}} \frac{Q^2}{(-t)}\cdot \Theta\left( \tilde{z} ,{\bf k}_{T1}^2, Q^2 \right), \label{eq:ST-t}
  \end{equation}
  where $t=(q_1-k_2)^2=-\left[Q^2(\tilde{z}-z)+ z\tilde{z}\cdot {\bf q}_{T1}({\bf q}_{T1}+2{\bf k}_{T1}) \right]/ z(1-\tilde{z})$ and the Transverse-Momentum Dependent splitting function $P_{QR}$ equals to:
  \begin{equation}
  P_{QR}(z,{\bf q}_{T1},{\bf q}_{T2})=T_R \frac{z^2({\bf q}_{T1}^2-2{\bf q}_{T1}{\bf q}_{T2})^2+2z({\bf q}_{T1}{\bf q}_{T2})({\bf q}_{T1}^2-2{\bf q}_{T1}{\bf q}_{T2}) + {\bf q}_{T1}^2 {\bf q}_{T2}^2}{{\bf q}_{T1}^2 \left( {\bf q}_{T2}^2 + z ({\bf q}_{T1}^2-2{\bf q}_{T1}{\bf q}_{T2}) \right)}. \label{eq:P_QR}
  \end{equation}
  
  Eq.~\ref{eq:P_QR} coincides with Eq.~13 of the Ref.~\cite{HHJ-Z} and reproduces the $P_{qg}(z)$ DGLAP splitting function in the collinear limit.
  
   The $u$-channel DCS term can be obtained from the $t$-channel DCS term (\ref{eq:ST-t}) via the substitution: $\tilde{z}\to 1-(\tilde{z}-z)$. Finally, subtracted HSC for the subprocess (\ref{eq:NLO-subp}) takes the form:
   \begin{equation}
    C_{2}^{(1,g)}\left( z, \frac{t_1}{Q^2} \right) = \int\limits_z^1  d\tilde{z} \int\limits_0^{2\pi}  \frac{d\psi}{2\pi} \left\{\frac{1}{(1-z)} {\cal C}^{(1,g)}_{2} -  \frac{1}{(1-\tilde{z})} \left. \Delta {\cal C}^{(t)}_{2}\right\vert_{\phi=\psi}^{\rm (mMRK)} - \frac{1}{(\tilde{z}-z)}  \left. \Delta {\cal C}^{(u)}_{2}\right\vert_{\phi=\psi}^{\rm (mMRK)}  \right\}. \label{eq:NLO-HSC}
   \end{equation}
   
   The analysis fully analogous to that of the Sec.~\ref{sec-2} shows, that if the HSC (\ref{eq:NLO-HSC}) is finite in collinear limit ${\bf q}_{T1}\to 0$, then the $O(\alpha_s)$ contribution of the subprocess (\ref{eq:NLO-subp}) is given by the convolution of the gluon PDF with the collinear limit of Eq.~\ref{eq:NLO-HSC}. Indeed, NLO HSC (\ref{eq:NLO-HSC}) is finite in this limit, despite the fact, that ${\cal C}^{(1,g)}_{2}$ contains collinear singularities for $t\to 0$ and $u\to 0$ when ${\bf q}_{T1}=0$. These singularities manifest themselves as $1/(\tilde{z}-z)$ and $1/(1-\tilde{z})$ poles in ${\cal C}^{(1,g)}_{2}$, but $\Delta {\cal C}^{(t)}_{2}$ and $\Delta {\cal C}^{(u)}_{2}$ also contain this poles when ${\bf q}_{T1}=0$, so that the integrand of (\ref{eq:NLO-HSC}) is free from any singularities even in the collinear limit. 
   
   It is convenient to study the collinear limit for Eq.~\ref{eq:NLO-HSC}, using dimensional regularization, since in this limit the unsubtracted NLO HSC in PRA coincides with the textbook result~\cite{Collins} for the unsubtracted NLO HSC of the subprocess (\ref{subp:gam_g-q_q}):
   \begin{equation}
  C_{2\ NS}^{(1,g)}(z,0)= C_{2g\ NS}^{(NLO\ CPM)}(z)=2\left[\frac{\bar{\alpha}_s(Q^2)}{2\pi} \left( - \frac{1}{\epsilon} - \log \frac{\mu^2}{Q^2}\right) P_{qg}(z) + C_{2g}^{\overline{MS}}(z)  + O(\epsilon)  \right], \label{eq:Cg_NS}
   \end{equation}    
   where $\epsilon=(4-D)/2$, $D$ is the dimension of space-time,   $\bar{\alpha}_s=(4\pi/\mu^2)^\epsilon e^{-\epsilon\gamma_E} \alpha_s $ is the dimensionless coupling constant of QCD in the $\overline{MS}$-scheme and $\gamma_E\simeq 0.5772$ is the Euler-Mascheroni constant. 
   
   One can derive the collinear limit of $t$-channel DCS term in $D$ dimensions, introducing the mMRK approximation for the partonic tensor into the standard calculation of the contribution of subprocess (\ref{subp:gam_g-q_q}) in CPM. The result reads:
   \begin{equation}
   \Delta C_2^{(t)}(z,0) = \frac{\bar{\alpha}_s(Q^2)}{2\pi} \left( \frac{\mu^2 z}{Q^2} \right)^\epsilon \int\limits_z^1 \frac{d\tilde{z}}{(1-\epsilon)} \frac{P_{qg}(\tilde{z},\epsilon)}{(\tilde{z}-z)^{1+\epsilon}} \cdot \Theta\left( \tilde{z}, Q^2\frac{\tilde{z}-z}{z},Q^2 \right) , \label{eq:DC_t-CL}
   \end{equation}
   where the factor $(1-\epsilon)$ in the denomiantor stands for averaging over $D-2=2(1-\epsilon)$ polarizations of the on-shell gluon in the initial state, and $P_{qg}(z,\epsilon)=T_R\left[ z^2+(1-z)^2 - \epsilon \right]$ is the space-like DGLAP splitting function in $D$-dimensions, see e.g. Eq.~4.15 in Ref.~\cite{CS-dipole}. The function $\Theta$, which implements KMR kinematic constraint~(\ref{eq:KMR-cut}), leads to the upper bound on $\tilde{z}<\tilde{Z}(z)$, where $\tilde{Z}(z)<1$ is the solution of a cubic equation:
   \begin{eqnarray}
  & \tilde{Z}(z)=\frac{1}{3}\left[ d^{1/3} - 2z(3-2z) d^{-1/3} + 2z \right],\ \  d=\frac{z}{2}\left[ 4z(4z-9) + 3\left(9+\sqrt{81+24z(2z-5)}\right) \right].\ \ \ \ \ 
   \end{eqnarray}    
   
   Calculating integral (\ref{eq:DC_t-CL}) and expanding the result in $\epsilon$ one obtains 
   \begin{equation}
  \Delta C_2^{(u)}(z,0) =  \Delta C_2^{(t)}(z,0) = \frac{\bar{\alpha}_s(Q^2)}{2\pi}\left(-\frac{1}{\epsilon}-\log\frac{\mu^2}{Q^2}\right) P_{qg}(z) + \Delta C_{qg}(z) + O(\epsilon), \label{eq:DC_CL}   
   \end{equation}
   where 
\begin{equation}
\Delta C_{qg}(z)= \frac{\alpha_s(Q^2)}{2\pi} \left\{\log\left(\frac{\tilde{Z}(z)-z}{z}\right)P_{qg}(z)+T_R\left[ \tilde{Z}^2(z)-2(1-z)\tilde{Z}(z)+z(4-5z)\right] \right\} .\label{eq:DC_qg} 
\end{equation}   

    Comparing Eqns.~\ref{eq:DC_CL} and \ref{eq:Cg_NS} one can see, that collinear divergences cancel explicitly, but the collinear limit of the subtracted HSC in PRA differs from the CPM result in $\overline{MS}$-scheme by the $\Delta C_{qg}(z)$ term:
   \begin{equation}
    C_{2}^{(1,g)}(z,0) = 2\left[ C_{2g}^{\overline{MS}}(z) - \Delta C_{qg}(z) \right]. 
   \end{equation}

\section{Numerical results}
\label{sec-4}

\begin{figure}
\begin{center}
\includegraphics[width=0.47\textwidth]{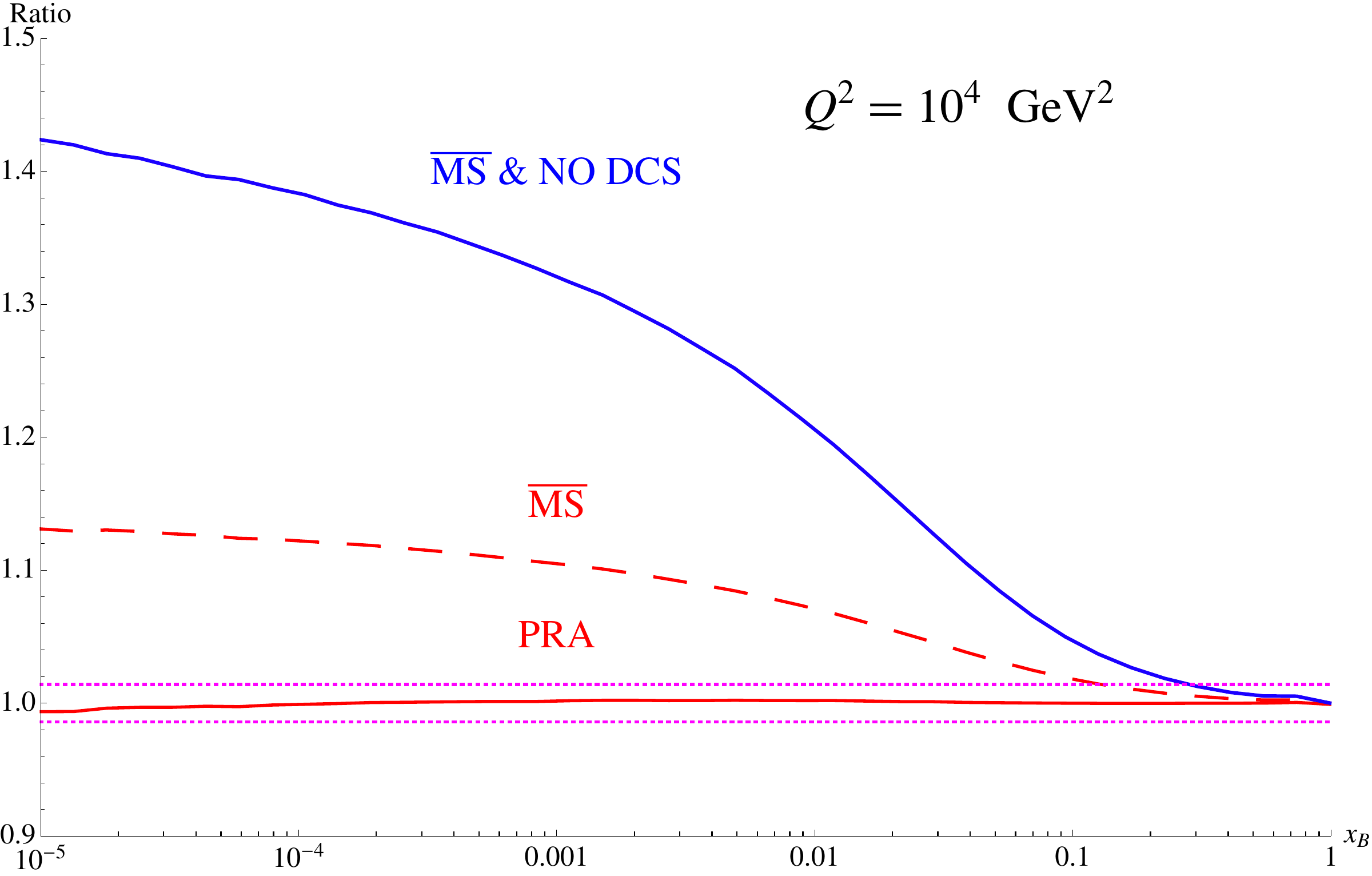}\ \ \  \includegraphics[width=0.47\textwidth]{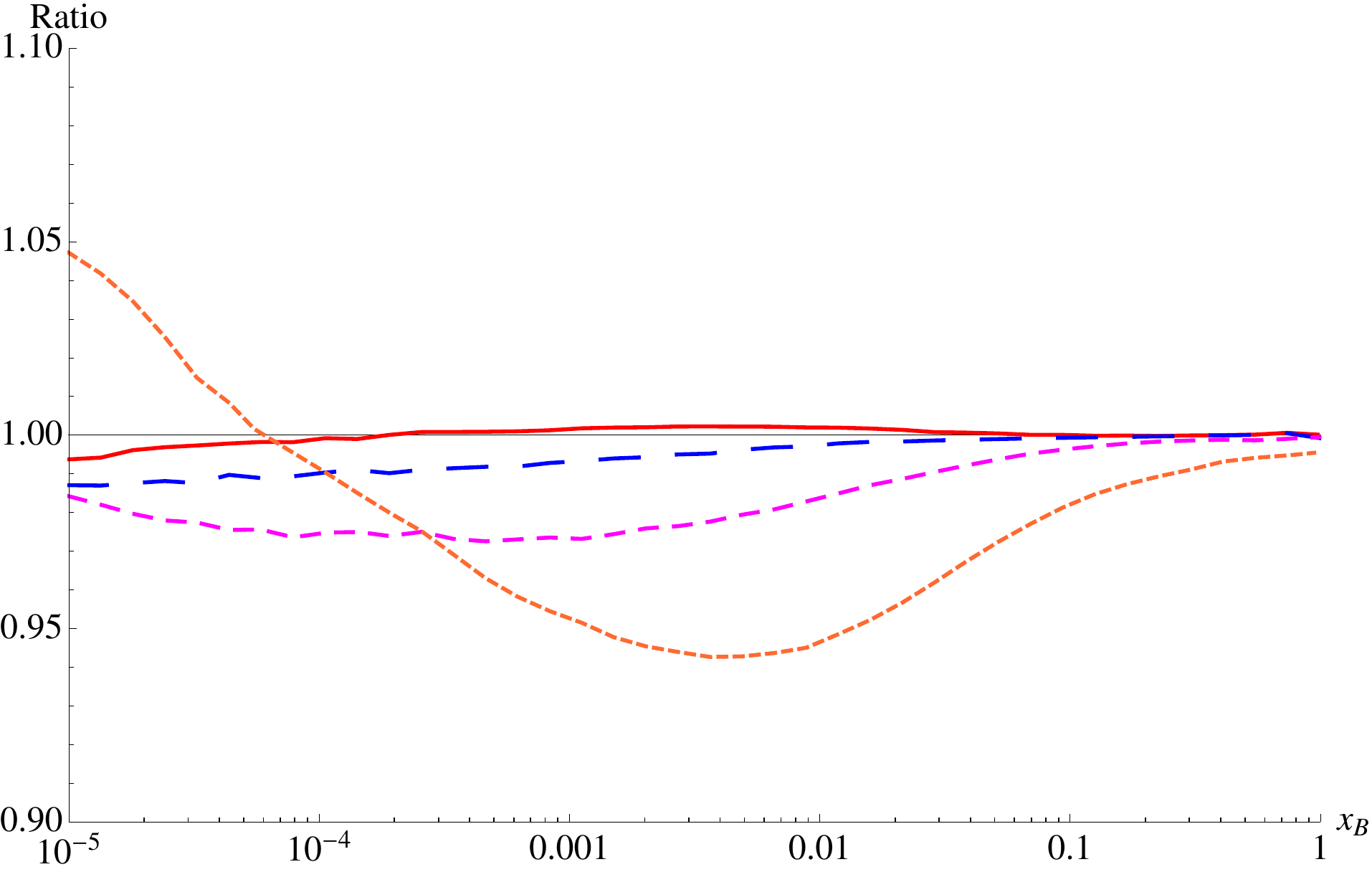}
\end{center}
\caption{Numerical results for the ratio of $\tilde{F}_{2q}^{(NLO\ PRA)}(x_B,Q^2)/\tilde{F}_{2q}^{(NLO\ CPM)}(x_B,Q^2)$ as a function of $x_B$. {\bf Left panel:} significance of different contributions to $\tilde{F}_{2q}^{(NLO\ PRA)}$ is illustrated. Curves from bottom to top: full PRA case;  $\overline{MS}\to PRA$ scheme transformation is {\it turned off}; subtraction of double-counting is {\it turned off}. Dotted horizontal lines denote $1\pm \alpha_s^2(Q^2)$ band. {\bf Right panel:} scale-dependence of the ratio. Solid line -- $Q^2=10^4$ GeV$^2$, dashed line -- $Q^2=10^3$ GeV$^2$, short-dashed line -- $Q^2=10^2$ GeV$^2$, dotted line -- $Q^2=10$ GeV$^2$.\label{fig:num}}
\end{figure}
 
  The PRA analog of CPM test-function (\ref{eq:CPM-test}) reads:
  \begin{eqnarray}
 \hspace{-0.6cm}\tilde{F}_{2q}^{(NLO\ PRA)}(x_B,Q^2) &=& (x_Be_q^2) \left\{ f^{PRA}_q(x_B,Q^2) + f^{PRA}_{\bar{q}}(x_B,Q^2) + 2 \left[\Delta C_{qg}^{(k_T)} \otimes f^{PRA}_g\right](x_B,Q^2) \right. \nonumber \\
 &+& \left. F_{2q}^{(1,g)}(x_B,Q^2)  \right\} , \label{eq:PRA-test-kT}
  \end{eqnarray}
  where $f_i^{PRA}$ are PDFs in the {\it PRA scheme}, precise definition of which is given in Eq.~\ref{eq:PRA-scheme} below. In the first line of Eq.~\ref{eq:PRA-test-kT}, the $O(\alpha_s^0)$ and $O(\alpha_s)$ terms from the LO PRA contribution (\ref{eq:factor}), relevant for our present analysis, are written. Term in the second line of Eq.~\ref{eq:PRA-test-kT} is given by Eq.~\ref{eq:F2q_1g}, i. e. it takes into account the ${\bf q}_{T1}$-dependence of NLO HSC. Using the results of Sec.~\ref{sec-3}, one can collect all $O(\alpha_s)$ contributions in Eq.~\ref{eq:PRA-test-kT} together:
  \begin{eqnarray}
 \hspace{-0.6cm}\tilde{F}_{2q}^{(NLO\ PRA)}(x_B,Q^2) &=&  (x_Be_q^2) \left\{ f^{PRA}_q(x_B,Q^2) + f^{PRA}_{\bar{q}}(x_B,Q^2) \right. \nonumber \\
 &+&\left. 2 \left[\left(C_{2g}^{\overline{MS}} - \Delta C_{qg}  + \Delta C_{qg}^{(k_T)} \right)  \otimes f^{PRA}_g\right](x_B,Q^2) + O(\alpha_s^2) + O\left( (Q^2)^{-\gamma} \right) \right\}, \label{eq:PRA-test-CL}
  \end{eqnarray}
  where $\gamma>0$. To ensure matching of the $O(\alpha_s^0)$ and $O(\alpha_s)$ terms in the Eq.~\ref{eq:PRA-test-CL} to the CPM result (\ref{eq:CPM-test}), PDFs in PRA scheme should be related with PDFs in $\overline{MS}$ scheme as follows:
  \begin{equation}
  f_g^{PRA}(x,\mu^2) = f_g^{\overline{MS}}(x,\mu^2),\ \ f_q^{PRA}(x,\mu^2) = f_q^{\overline{MS}}(x,\mu^2) + \left[ \left( \Delta C_{qg} - \Delta C_{qg}^{(k_T)} \right)  \otimes f_g^{\overline{MS}} \right] (x,\mu^2). \label{eq:PRA-scheme}
  \end{equation}
  
  In the complete NLO PRA calculation, all LO and NLO terms should be written in $k_T$-factorized form, {\it with the unPDFs normalized to the NLO PDFs in PRA scheme as prescribed by the Eq.~\ref{eq:unPDF-norm}}. To this end, the usual $\overline{MS}$ PDFs shold be transformed into PRA scheme by the transformation, similar to the transformation (\ref{eq:PRA-scheme}), and then the unPDFs can be constructed from them, using Eqns.~\ref{eq:unPDF-int} and \ref{eq:Sudakov}. NLO splitting functions in this equations also recieve the $O(\alpha_s)$ corrections, due to the difference between $\overline{MS}$ and $PRA$ schemes. These corrections are known, but the corresponding formulae are too lengthy to present them here. In contrast to the complete NLO calculation, in the present paper we have concentrated on the contribution of only one subprocess (\ref{eq:NLO-subp}), and therefore, PDFs in PRA scheme appear explicitly in Eq.~\ref{eq:PRA-test-kT}.

  In the Fig.~\ref{fig:num} the PRA test function (\ref{eq:PRA-test-kT}) is compared numerically to the CPM test function (\ref{eq:CPM-test}). The NLO set of MSTW-2008 PDFs~\cite{MSTW-2008} has been used as the collinear input for this computation, togeher with the value of $\alpha_s(M_Z)=0.1202$ from the PDF fit.  As it is explained above, Eq.~\ref{eq:PRA-test-kT} should reproduce the CPM result up to $O(\alpha_s^2)$ terms and corrections supressed by the negative power of $Q^2$. Numerically, NLO PRA result (\ref{eq:PRA-test-kT}) and NLO CPM result (\ref{eq:CPM-test}) indeed agree with the accuracy better than $1\%$ for $Q^2=10^4$ GeV$^2$ (see the right panel of Fig.~\ref{fig:num}). Their agreement become less good with $Q^2$ decreasing, but even for the $Q^2=10$ GeV$^2$ the error does not exceed $\pm 7\%$, which is compatible with the value of $\alpha_s^2(Q^2)$ at this scale. 
  
  In the left panel of Fig.~\ref{fig:num}, the role of different contributions in the PRA result is illustrated. Neglecting the scheme-transformation (\ref{eq:PRA-scheme}) one clearly misses some $O(\alpha_s)\sim O(10\%)$ correction, while turning off the DCS terms one overshoots the NLO CPM result by $40\%$ in the small-$x_B$ region.     
  
  In the present contribution, the problem of double-counting of real NLO corrections in PRA is adressed, and the procedure for the subtraction of this double counting, together with the definition of NLO unPDF in PRA is constructed, which resolves this problem and correctly takes into account the scheme-dependence of NLO PDFs in CPM. In such a way, significant part of the formalism of NLO calculations in PRA has been formulated.  
  
  {\bf Acknowledgements:}  This work was supported by the Ministry of Education and Science of Russia under Competitiveness Enhancement Program of Samara
University for 2013-2020, project 3.5093.2017/8.9.


\begin{thebibliography}{}
%
%
\bibitem{Loop_proc}  M.~Nefedov and V.~Saleev,
  ``Towards NLO calculations in the parton Reggeization approach,''
  arXiv:1608.04201 [hep-ph].
  
\bibitem{IFL}
B. L. Ioffe, V. S. Fadin and L. N. Lipatov, \textit{Quantum chromodynamics, perturbative and nonperturbative aspects} (Cambridge University Press, Cambridge, UK, 2011)

\bibitem{Collins} J. C. Collins, {\it Foundations of preturbative QCD} (Cambridge University Press, Cambridge, UK, 2011)

\bibitem{Proc_HQ}  A.~Karpishkov, M.~Nefedov, V.~Saleev and A.~Shipilova,
  ``Heavy quark production at the LHC in the Parton Reggeization Approach,''
  arXiv:1702.05081 [hep-ph].
  
\bibitem{BB-correlations}  A.~Karpishkov, M.~Nefedov and V.~Saleev,
  ``$B{\bar B}$ angular correlations at the LHC in parton Reggeization approach merged with higher-order matrix elements,''
  arXiv:1707.04068 [hep-ph].
  
\bibitem{Fadin-Sherman} V. S. Fadin and V. E. Sherman, JETP Lett. {\bf 23}, 599 (1976); JETP {\bf 45}, 861 (1977).

\bibitem{ReggeProofLL}  A.~V.~Bogdan and V.~S.~Fadin,
 ``A Proof of the reggeized form of amplitudes with quark exchanges,''
  Nucl.\ Phys.\ B {\bf 740}, 36 (2006). 

\bibitem{KMR}   M.~A.~Kimber, A.~D.~Martin and M.~G.~Ryskin, ``Unintegrated parton distributions,'' Phys.\ Rev.\ D {\bf 63}, 114027 (2001);  A.~D.~Martin, M.~G.~Ryskin and G.~Watt,
  ``NLO prescription for unintegrated parton distributions,''
  Eur.\ Phys.\ J.\ C {\bf 66}, 163 (2010).

\bibitem{MC-rev} A.~Buckley {\it et al.},
  ``General-purpose event generators for LHC physics,''
  Phys.\ Rept.\  {\bf 504}, 145 (2011).


\bibitem{HHJ-Z} F.~Hautmann, M.~Hentschinski and H.~Jung,
  ``Forward Z-boson production and the unintegrated sea quark density,''
  Nucl.\ Phys.\ B {\bf 865}, 54 (2012).

\bibitem{CS-dipole}  S.~Catani, S.~Dittmaier, M.~H.~Seymour and Z.~Trocsanyi,
  ``The Dipole formalism for next-to-leading order QCD calculations with massive partons,''
  Nucl.\ Phys.\ B {\bf 627}, 189 (2002).

\bibitem{MSTW-2008}  A.~D.~Martin, W.~J.~Stirling, R.~S.~Thorne and G.~Watt,
  ``Parton distributions for the LHC,''
  Eur.\ Phys.\ J.\ C {\bf 63}, 189 (2009).
 
\end{thebibliography}
%
%

\end{document}